\documentclass[12pt]{article}

\usepackage{newtxtext,newtxmath}

\usepackage{graphicx}
\usepackage{tabularx}

\usepackage[letterpaper,margin=1in]{geometry}

\renewenvironment{abstract}
	{\quotation}
	{\endquotation}

\date{}

\makeatletter
\renewcommand{\fnum@figure}{\textbf{Figure \thefigure}}
\renewcommand{\fnum@table}{\textbf{Table \thetable}}
\makeatother

\usepackage{scicite}
\usepackage{comment}

\usepackage{url}

\def\scititle{
	Liquid combs: broadband light with equidistance and without stability
}
\title{\bfseries \boldmath \scititle}

\author{
	Mithun~Roy$^{1}$,
	Tianyi~Zeng$^{2,3}$,
    Zhenyang~Xiao$^{1}$,
    Chao~Dong$^{1}$,\and
    Sadhvikas~Addamane$^{4}$,
    Qing~Hu$^{3}$,
	David~Burghoff$^{1\ast}$\and
	\small$^{1}$Chandra Department of Electrical and Computer Engineering, Cockrell School of Engineering,\and
     \small University of Texas at Austin, Austin, Texas 78712, USA.\and
    \small$^{2}$John A. Paulson School of Engineering and Applied Science, Harvard University,\and
    \small Cambridge, Massachusetts 02138, USA.\and
    \small$^{3}$Department of Electrical Engineering and Research Laboratory of Electronics,\and
    \small Massachusetts Institute of Technology, Cambridge, Massachusetts 02139, USA.\and
	\small$^{4}$Center for Integrated Nanotechnologies, Sandia National Laboratories,\and
    \small Albuquerque, New Mexico 87185, USA.\and
	\small$^\ast$Corresponding author. Email: burghoff@utexas.edu
}

\begin{document} 

\maketitle

\begin{abstract} \bfseries \boldmath
Broadband light sources with well-defined spectral structures are vital for science and technology. However, the evenly spaced lines of frequency combs represent only a small subset of all possible structured white-light sources. We demonstrate liquid combs: optical states that preserve spectral equidistance but lack temporal stability.  By engineering the gain and dispersion of semiconductor laser cavities, we produce light that possesses rapid phase fluctuations but maintains relative phase differences between modes that vary identically. We show experimentally that this phenomenon occurs in multiple laser platforms—across multiple octaves—through the creation of a metrological technique that determines the phase differences. We also show theoretically that this is a general phenomenon that can be described using a mean-field theory. These liquid combs are attractive for many applications due to having wider bandwidths than frequency combs, and more generally, they represent the long-sought realization of structured white light sources that are not combs.

\end{abstract}



\noindent
Sources with broad bandwidths and well-defined, discrete spectra are essential for progress in spectroscopy, metrology, quantum optics, and communication. Frequency combs---sources with equidistant and stable modes---stand out as a prominent example \cite{suh_microresonator_2016,yu_silicon-chip-based_2018,spencer_optical-frequency_2018,trocha_ultrafast_2018,suh_soliton_2018,brasch_photonic_2016,delhaye_phase-coherent_2016}. 
Although stability is useful, their main efficacy actually derives from their mutual coherence and equidistance. It is equidistance that makes them useful for applications like dual-comb spectroscopy, since equidistance allows the position of one line to be determined from another using a low-frequency measurement. But actually, for many applications, the comb structure is not needed at all! Any spectral structure could be used, \textit{provided that the structure is known}. The primary issue is that, to date, no broadband light source with a well-defined spectral structure has been created that is \textit{not} a comb.



\begin{figure} 
	\centering
	\includegraphics{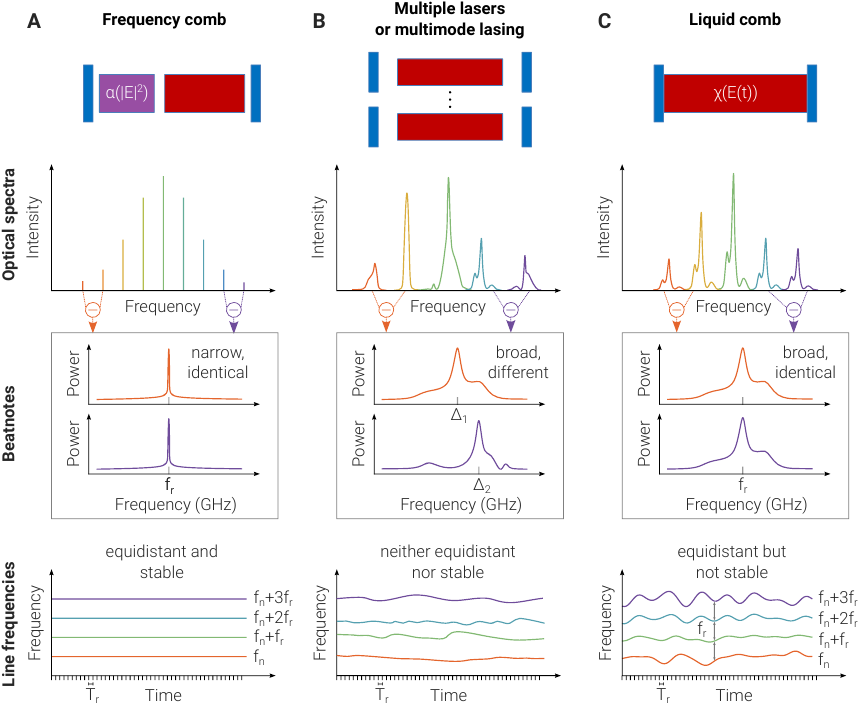} 
        \caption{\textbf{Comparison of different broadband light sources.}
Emission spectra, intermodal beat spectra, and frequencies of the constituent modes for \textbf{(A)} a typical frequency comb, \textbf{(B)} a multi‐laser or multimode system, and \textbf{(C)} a liquid comb.
The spectrum of a frequency comb consists of equidistant lines, producing narrow beatnotes with a common center frequency. Consequently, the modal phases vary linearly with time.
In a multi‐laser or multimode laser, the lack of mutual coherence broadens individual beatnotes and misaligns their center frequencies, and there is no correlation among the nonlinear phase components.
By contrast, the modes of a liquid comb exhibit uniform phase variation over timescales of thousands of round trips, leading to identical frequency fluctuations. In other words, the phase differences between the modes remain identical.}
\label{fig_concept}
\end{figure}

The two salient properties of frequency combs are illustrated in Fig. \ref{fig_concept}A. Not only are the 
frequencies of comb lines equidistant, but those frequencies are also stable over many round trips. When beating any pair of lines together, a single, narrow beatnote is produced. Moreover, measurement of this beatnote allows one to determine the positions of all lines from the position of just one: this is the power of a comb. The first and most common combs were generated by passive modelocking, but semiconductor lasers can also spontaneously produce frequency combs using an internal mode-locking mechanism. Those combs are predominantly frequency-modulated (FM) in character and have quasi-continuous-wave temporal intensity and a distinct, often linear, frequency chirp \cite{singleton_evidence_2018,hillbrand_-phase_2020,schwarz_monolithic_2019,sterczewski_frequency-modulated_2020, day_simple_2020, cappelli_retrieval_2019, senica_frequency-modulated_2023}. This is in stark contrast to an array of lasers or a multimode laser, which produces a series of lines that are neither equidistant nor stable (Fig. \ref{fig_concept}B).

However, semiconductor lasers engineered for comb operation frequently enter multimode states that appear to possess low mutual coherence, especially under high-bias conditions \cite{schwarz_monolithic_2019,sterczewski_interband_2021,villares_dispersion_2016, mezzapesa_tunable_2019}. This regime is identified by a \textit{broad} intermodal beat spectrum, with a full width at half maximum (FWHM) typically in the MHz range, corresponding to fluctuations over hundreds of round trips.
An FM comb requires dispersion above a minimum value 
to remain stable, which effectively limits the achievable bandwidth\cite{opacak_theory_2019,burghoff_unraveling_2020, humbard_analytical_2022,roy_fundamental_2024}. When the dispersion falls below this minimum value, the laser can transition into a broad-beatnote multimode regime. Interestingly, due to their low dispersion, these states often have more uniform spectra and wider bandwidths compared to the comb states \cite{rosch_octave-spanning_2015}. Such a state is often referred to as a pseudorandom state \cite{khurgin_coherent_2014}, as in simulation, the parameters vary chaotically over timescales much longer than a repetition period but remain comb-like over short timescales. It is also connected to the Risken-Nummedal-Graham-Haken instability \cite{risken_selfpulsing_1968,wang_coherent_2007}. Intriguingly, these states are not merely the modes of a dispersed cavity: they have smooth spectra and highly repeatable beatnotes. They are often even more robust than the comb state. To date, most research efforts have focused on characterizing and improving traditional combs. Although the broad-linewidth regime frequently occurs in semiconductor lasers, it has received considerably less attention and remains poorly understood.  
In this work, we show for the first time that these spontaneously formed states can possess a high degree of mutual coherence, even when they lack temporal stability. 
For a narrow-linewidth comb, the phase differences between consecutive lines vary linearly with time. For these broad-beatnote states, while the variation is highly nonlinear, the phases evolve in \textit{unison}---that is, the frequencies remain perfectly evenly spaced (Fig. \ref{fig_concept}C).
We refer to these states as \textit{liquid combs} in analogy to liquid crystals, which can flow but maintain long-range orientational order \cite{misra_nonlinear_2012}. 
Liquid combs have fundamentally different emission when compared with an array of lasers or a multimode laser, as these exhibit no correlation in the modal phase differences due to a lack of mutual coherence (Fig. \ref{fig_concept}B).


This work was enabled by the realization of a metrology scheme suitable for the characterization of broad-beatnote devices. Novel metrologies are of critical importance, as the pace of innovation in metrology often lags behind new device discoveries. 
For example, although mid-infrared quantum cascade laser combs were first demonstrated in 2012 \cite{hugi_mid-infrared_2012}, the nature of their instantaneous intensity and frequency was not determined until 2018 \cite{singleton_evidence_2018}. Since these combs are usually not pulses, conventional methods relying on optical nonlinearities, such as FROG and interferometric autocorrelation, could not be used for their characterization. The Shifted Wave Interference Fourier Transform Spectroscopy (SWIFTS) technique \cite{burghoff_terahertz_2014,burghoff_evaluating_2015} remedied this by allowing low intensity combs to be characterized, and this technique has become widely used across lasers spanning the terahertz (THz) to the visible \cite{hillbrand_-phase_2020,sterczewski_frequency-modulated_2020,han_sensitivity_2020, heckelmann_quantum_2023-1}. 
SWIFTS confirmed that FM-comb formation is a general phenomenon, and other measurements \cite{taschler_asynchronous_2023,cappelli_retrieval_2019} subsequently verified these findings. 
However, while SWIFTS is suitable for low-intensity combs, it does not apply to spectra that are not combs. We develop a version of SWIFTS that is suitable not just for rigorously measuring the coherence of broad-beatnote states like liquid combs, but can even be applied to characterize arbitrary coherent spectra. This new approach enables the measurement of the coherence as a function of frequency, thereby confirming uniform phase variation in liquid combs.

Furthermore, in this work, we show experimentally that the liquid comb state is a general phenomenon in chip-scale lasers and is not specific to one frequency range. We demonstrate this phenomenon in both a mid-infrared (mid-IR) and a THz QCL. 
The mid-IR QCL is based on an InGaAs/AlInAs heterostructure with a dielectric waveguide, while the THz QCL is based on GaAs/AlGaAs and was fabricated using a metal-metal waveguide. Despite the order of magnitude difference in frequency, the different material systems, and the different types of waveguides, both lasers exhibit regimes of liquid combs. Since these comb states are not subject to the same limits as FM combs, they frequently have bandwidths that are far greater than their comb counterparts, making them even better for most applications where combs are used. Furthermore, since liquid combs exhibit uniform phase variations, it should be possible to correct their nonlinear phases either experimentally \cite{singleton_pulses_2019} or computationally \cite{burghoff_computational_2016}.



\subsection*{\label{results}Results}

\begin{figure} 
	\centering
	\includegraphics{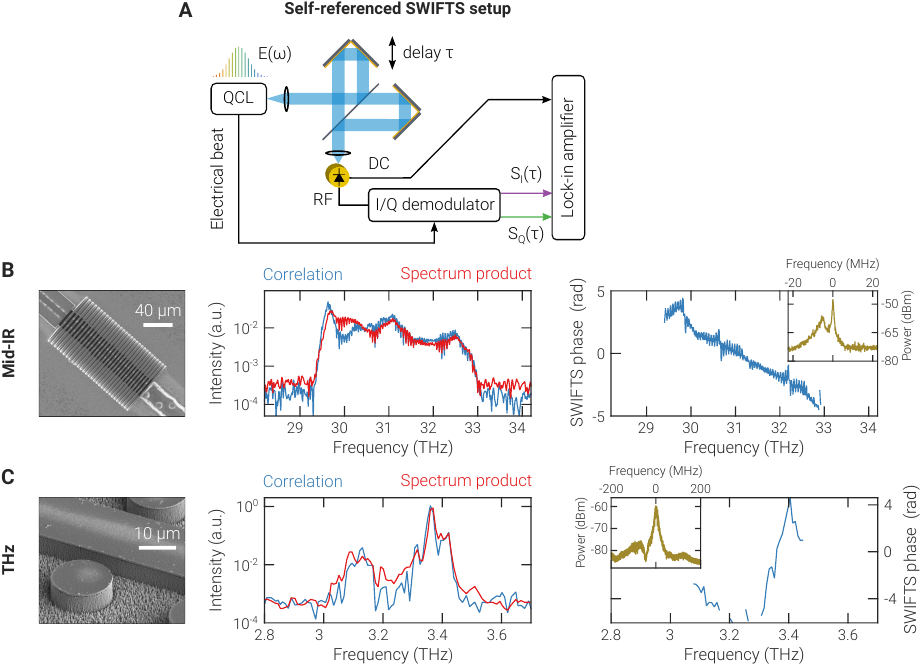} \caption{\textbf{Self-referenced SWIFTS of liquid combs.}  
    \textbf{(A)} Schematic diagram of the setup for self-referenced SWIFTS, in which the electrical betanote is used as the reference frequency for the in-phase-and-quadrature demodulator. 
 Scanning-electron-microscope images, correlation spectra and spectrum products, and phases of the correlation spectra (SWIFTS phases) for the \textbf{(B)} mid-IR and \textbf{(C)} THz QCLs used in our work. 
 The mid-IR QCL employs a double-chirped mirror, while the THz QCL employs small disks on both sides of the Fabry-Perot cavity. 
 The FWHM values of the intermodal radio-frequency beatnotes, shown in the inset, are $\sim\!0.41$ and $\sim\!12.6$ MHz, respectively.
}
\label{fig_swifts_self}
\end{figure}

In order to prove that the formation of liquid combs is a general phenomenon, both a THz and a mid-IR QCL were fabricated and engineered for gain curvature and low group velocity dispersion, respectively. 
The mid-IR QCL 
contains an integrated double-chirped mirror for dispersion engineering (Fig. \ref{fig_swifts_self}B, left panel). 
For the THz QCL,
small disks (pillars), composed of the active-region material and enclosed by a top and bottom metal layer, were added to alter the curvature of the laser gain.
The disks were positioned on both sides of the Fabry-Perot cavity, covering nearly the entire cavity length (Fig. \ref{fig_swifts_self}C, left panel). For specific parameter choices (such as the radius and coupling distance), the disks induce light coupling and introduce resonant loss \cite{roy_fundamental_2024}. The parameters were chosen such that the center frequency of the resonant loss coincides with that of the unsaturated gain of the QCL, thus modifying the gain curvature (see ``Device design and measurements" in Supplementary Materials).

The mid-IR QCL showed both narrow and broad-beatnote regimes---a typical behavior caused by the bias-current-induced change in dispersion \cite{burghoff_terahertz_2014, lu_dispersion_2017}.
For the THz QCL, the peak value of the added resonant loss was high ($\sim\!8$ cm$^{-1}$). Consequently, the laser exhibited only a broad beatnote regime (see Fig. S2A). 
To approximately assess the g$^{(1)}$-coherence of the broad-linewidth states, we performed self-referenced SWIFTS,
where the electrical beatnote was used as the reference frequency (Fig. \ref{fig_swifts_self}A). The basic idea of SWIFTS is to use an interferometer to demultiplex the many pairs of beatings contributing to a comb's beatnote; once these beatings are isolated, one can compare how much each adjacent line is beating to the optical spectrum. For true combs, one can only measure a non-zero SWIFTS signal when the reference frequency perfectly matches the comb spacing, and phase-locking or injection locking is required to match the comb to the reference. In the self-referenced version, the beatnote is instead derived from the laser itself to facilitate this (see ``Self-referenced SWIFTS" in Supplementary Materials).

The striking result of this measurement is that the self-referenced SWIFTS implies that these states are actually extremely coherent,
provided that one generalizes the notion of coherence to allow the reference frequency itself to be fluctuating in time. For the mid-IR laser, there is a good agreement between the spectrum product derived from the normal interferogram and the SWIFTS correlation spectrum (Fig. \ref{fig_swifts_self}B, center panel).
For the THz QCL, a trough is present in the spectra at $\sim\!3.23$ THz (Fig. \ref{fig_swifts_self}C, center panel), caused by the resonant loss introduced by the disks. It is worth noting that two different detectors---a Schottky mixer for the SWIFTS correlation and a highly sensitive bolometer for the spectrum product---were used, and they were not operated simultaneously. Despite this, the spectrum product showed good agreement with the correlation spectrum. The discrepancy in the spectral trough is attributed to the difference in the sensitivity of the two detectors. 
Moreover, the SWIFTS phase for the mid-IR QCL is approximately linear (linear chirp in the instantaneous frequency), whereas the THz QCL did not show the same behavior (Figs. \ref{fig_swifts_self}B and \ref{fig_swifts_self}C, right panels). This difference can be attributed to the much larger gain bandwidth of the mid-IR QCL and faster gain dynamics \cite{silvestri_unified_2024-1}. Still, for both lasers, these results are strongly suggestive of a tantalizing possibility: even though both states have a broad beatnote, which usually indicates incoherence, each of the spectral teeth is, in fact, separated by \textit{the same} unstable frequency. In other words, the phase of each component can be written in terms of its neighbor as $\phi_{i+1}(t)-\phi_i(t)=\Delta_{i,i+1} +\phi_r(t)$, where $\phi_r(t)$ is a nonlinear difference phase that gives rise to the wide beatnote, and $\Delta_{i,i+1}$ is a constant.



\begin{figure} 
	\centering
	\includegraphics{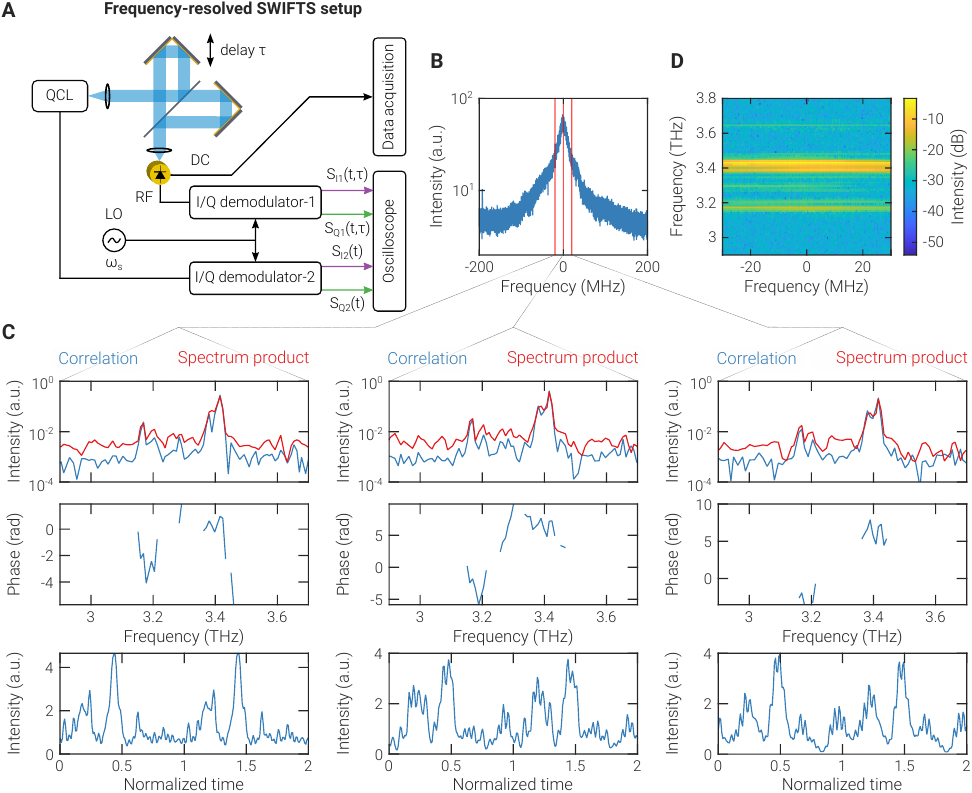}\caption{\textbf{Frequency-resolved SWIFTS measurement of liquid combs.}  
\textbf{(A)} Schematic diagram of the setup. A radio-frequency synthesizer is used to downconvert both the electrical and optical beatnotes, and a high-bandwidth oscilloscope is used to collect the downconverted in-phase and quadrature outputs. 
\textbf{(B)} Downconverted optical beatnote from the THz QCL at a specific interferometer delay. The QCL was biased at 0.2~A. The FWHM of the beatnote is $\sim\!12.5$ MHz.
\textbf{(C)} Correlation spectra and spectrum products, phases of the correlation spectra, and temporal intensities at the downconverted frequencies of $-20$, $0$, and $20$ MHz [indicated by the red lines in (B)]. A Schottky detector was used for both the correlation spectrum and the spectrum product. The agreement between the two indicates uniformity in the temporal phase differences between the modes. 
\textbf{(D)} Spectrogram of the correlation spectrum as a function of beatnote frequencies.
(See Fig. S3 for the measurement of a liquid comb at a bias current of 0.24 A.)
}
\label{fig_swifts_freq}
\end{figure}

However, self-referenced SWIFTS is not a fully rigorous coherence measurement. One could imagine, for example, that different parts of the spectrum create different components of the broad beatnote. In this case, one would still obtain a non-zero signal at each optical frequency since the beatnote sweeps across a broad range of radio frequencies and the final result is averaged. One would then get a coherence spectrum that resembles the original spectrum product, but is smaller by a factor of the number of comb teeth (i.e, $g^{(1)}=1/N$ instead of 1). To conclusively prove coherence, we devised a frequency-resolved SWIFTS measurement that can evaluate coherence at all beatnote frequencies. In a conventional SWIFTS setting, one measures the optical signal using a fast detector ($V(t,\tau)$) as a function of time t and delay $\tau$. Then, one considers $ \langle V(t,\tau)\, \mathrm{e}^{-\mathrm{i} \omega_\mathrm{r} t} \rangle $, where $ \omega_\mathrm{r} $ is the reference frequency—typically derived from the electrical beatnote—and the angle brackets denote an average over time $ t $. As a result, only the Fourier component $ \tilde{V}(\omega_\mathrm{r},\tau) $ is retained. 
For SWIFTS done on broad-beatnote states in the self-referenced scheme, $\omega_\mathrm{r}$ consists of a range of frequencies. Therefore, to rigorously evaluate coherence, one must analyze the individual Fourier components \(\tilde{V}(\omega,\tau)\). 
Hence, we devised a setup that allows access to these Fourier components (Fig. \ref{fig_swifts_freq}A). 
In particular, both the optical and electrical beatnotes are downconverted and collected using a high-bandwidth oscilloscope (see ``Frequency-resolved SWIFTS" in Supplementary Materials). 
The downconverted optical beatnote for the THz QCL at a specific interferometric delay is shown in Fig. \ref{fig_swifts_freq}B, indicating a 10-dB signal-to-noise ratio and an FWHM of $\sim\!12.5$ MHz. 
Figure \ref{fig_swifts_freq}C shows the coherence and temporal properties for three different frequencies ($\omega$). A Schottky detector was used for both the spectrum product and the correlation spectrum. Despite the detector's limited signal-to-noise ratio in its DC-coupled monitor port, the two spectra agreed well in all cases. 
The phase of the correlation spectrum and the temporal intensity are similar to those observed using the self-referenced scheme in Fig. \ref{fig_swifts_self}C and Fig. S1, showing both frequency and amplitude modulations.
A spectrogram of the correlation spectrum shows that it remains nearly unchanged over the whole range of frequencies around which the beatnote has a high signal-to-noise ratio (Fig. \ref{fig_swifts_freq}D).

What does the agreement between the correlation and spectrum product over a range of frequencies indicate? Using elementary Fourier analysis, one can express the frequency resolved SWIFTS as $ \tilde{V}(\omega,\tau)=\mathcal{F}_t\Bigl[\sum_{m} E_{m+1} E_m^* \allowbreak \, \mathrm{e}^{\mathrm{i}\,\Omega_m \tau} \, \allowbreak \mathrm{e}^{\mathrm{i}\, \phi_{m+1,m}(t)}\Bigr](\omega)$, where $\mathcal{F}_t$ denotes the Fourier transform with respect to $t\,$, $E_m$ and $\Omega_m$ are the amplitude and the frequency of the $m$th mode, and $\phi_{m+1,m}(t)$ is the phase difference between the corresponding modes \cite{burghoff_evaluating_2015}. Since this is mathematically identical to conventional SWIFTS, the same rigorous coherence bounds (based on the Cauchy-Schwarz inequality) apply, and one can only expect a coherence spectrum proportional to the spectrum product signal at a particular $\omega$ if that pair of spectral components is fully coherent. Therefore, the good agreement across all $\omega$ in the beatnote confirms that all modes remain \textit{equidistant}, even as the beatnote fluctuates. Moreover, all of the optical frequencies are present in all of the beatnote frequencies.
Thus, one can visualize a liquid comb as a state in which the repetition rate fluctuates over time while all the individual phase differences vary in an identical manner.


\subsection*{\label{discussions}Discussion}
What prompts semiconductor lasers to enter the liquid comb regime? Qualitatively, light in any gain medium with fast gain recovery dynamics (total or partial) has an intensity profile that behaves similarly to the surface of a liquid \cite{dikopoltsev_collective_2025}, as the fast recovery prefers a constant intensity profile. Quantitatively, this question can be addressed using a mean-field theory \cite{burghoff_unraveling_2020, burghoff_combs_2024} originally formulated by adapting the famed Lugiato-Lefever equation \cite{cole_theory_2018, lugiato_traveling_2018}
to active bidirectional cavities. 
The fundamental FM comb state---referred to as an extendon---exhibits narrow, stable beatnotes when its amplitude modulation is small. As the dispersion is reduced (for instance, by employing a dispersion-compensation mechanism), the extendon spans a broader bandwidth and consequently experiences higher amplitude modulation. Ultimately, when the dispersion falls below an optimum value, the extendon transitions into the broad-beatnote regime. Gain curvature can also play a role: under small values of gain curvature, extendons experience low amplitude modulation and thus form narrow beatnotes \cite{roy_fundamental_2024}, while higher values can push the extendon into the liquid comb regime, as is the case with the THz QCL in this work.

\begin{figure} 
	\centering
	\includegraphics{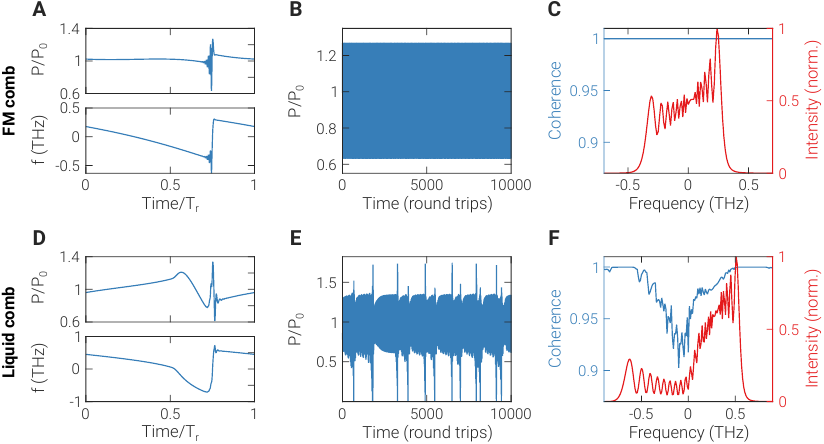} \caption{\textbf{Mean-field simulation of typical FM and liquid combs.}  
Instantaneous intensities and frequencies over a single round trip, instantaneous intensities over many round trips, and mutual coherence and spectra for an \textbf{(A)}--\textbf{(C)} FM comb and a \textbf{(D)}--\textbf{(F)} liquid comb. The intensities were normalized to the steady-state intensities at the left cavity facet. The group velocity dispersion was set to 1000 fs$^2$/mm for the FM comb and 582 fs$^2$/mm for the liquid comb.
}
\label{fig_simulation}
\end{figure}

To verify the above interpretation, a typical mid-IR QCL was simulated using the mean-field theory (see ``Mean-field simulation" in Supplementary Materials). 
The mutual coherence is defined as
\(
|\langle e^{\mathrm{i}\,\psi_m(t)}\rangle|,
\)
where the angle brackets represent a time average and \(\psi_m(t)\) is the second difference in phase between the corresponding modes. For a dispersion value higher than the optimum, an FM comb forms with nearly constant intensity 
(except at the point where the instantaneous frequency is discontinuous) and a linear chirp (Fig.~\ref{fig_simulation}A). The intensity remains stable for many round trips (Fig.~\ref{fig_simulation}B), and the mutual coherence becomes unity across all modes 
(Fig.~\ref{fig_simulation}C). However, when the dispersion falls below the optimum value, the extendon deviates from the 
quasi-continuous-wave and linear-chirp behavior of an FM comb (Fig.~\ref{fig_simulation}D). 
Moreover, the intensity no longer remains stable over many round trips, undergoing rapid fluctuations (Fig.~\ref{fig_simulation}E). Despite these intensity fluctuations, the mutual coherence remains \textit{high} (\(>\!0.9\)), indicating uniformity in the temporal phase differences between the modes (Fig.~\ref{fig_simulation}F).

In conclusion, we have demonstrated a distinct class of broadband optical states—liquid combs—that maintain strict spectral equidistance despite lacking temporal stability. These states are useful for many applications and can be an alternative to frequency combs. For example, their uniform phase variations should allow the use of computational methods to correct the complex multiheterodyne spectrum in dual-comb spectroscopy \cite{burghoff_computational_2016,sterczewski_computational_2019-1}, and with fast phase modulators and grating compressors, these states could even be converted into pulses\cite{taschler_femtosecond_2021,martinez_3000_1987}. Moreover, since they occur at lower dispersion values, liquid combs often possess broader bandwidths than traditional combs \cite{rosch_octave-spanning_2015}, which makes them highly suitable for spectroscopy and sensing applications. This newfound class of coherent broadband sources broadens the landscape of structured light, enabling more flexible designs for applications in spectroscopy, sensing, and communications. More generally, our work establishes liquid combs as a practical and versatile alternative to conventional frequency combs, offering new opportunities for the generation of broadband light with precise spectral structure.


\clearpage 

%
\bibliography{my_ref,my_ref2}
\bibliographystyle{sciencemag}

%
%
%
%
%
%


\section*{Acknowledgments}
\paragraph*{Funding:}
 D.B. acknowledges support from ONR grant N00014-21-1-2735,  AFOSR MURI grant no. FA9550-24-1-0349, and NSF grant ECCS-2046772; this research is funded in part by the Gordon and Betty Moore Foundation through Grant GBMF11446 to the University of Texas at Austin to support the work of D.B. This work was performed, in part, at the Center for Integrated Nanotechnologies, an Office of Science User Facility operated for the US Department of Energy (DOE) Office of Science. Sandia National Laboratories is a multimission laboratory managed and operated by National Technology \& Engineering Solutions of Sandia, LLC, a wholly-owned subsidiary of Honeywell International, Inc., for the US DOE’s National Nuclear Security Administration under contract DE-NA-0003525. The views expressed in the article do not necessarily represent the views of the US DOE or the United States Government.
 

\paragraph*{Competing interests:}
There are no competing interests to declare.







\newpage

\end{document}